\documentclass[aps,prd,preprint,superscriptaddress,showpacs,floatfix,nofootinbib]{revtex4}

\usepackage{color}

\def\bwt{\begin{widetext}}
\def\ewt{\end{widetext}}
\def\be{\begin{equation}}
\def\ee{\end{equation}}
\def\bea{\begin{eqnarray}}
\def\eea{\end{eqnarray}}
\def\bean{\begin{eqnarray*}}
\def\eean{\end{eqnarray*}}
\def\bary{\begin{array}}
\def\eary{\end{array}}
\def\bit{\begin{itemize}}
\def\eit{\end{itemize}}

\def\su5u1{SU(5) \times U(1)}
\def\fsu5u1{SU(5) \times U(1)'}
\def\so10{SO(10)}
\def\sq20{SO(10) \times SO(10)}

\begin{document}

\title{Distinguishing the right-handed up/charm quarks
from top quark via discrete symmetries in the 
standard model extensions }

\author{Chao-Shang Huang}
\affiliation{State Key Laboratory of Theoretical Physics,
 Institute of Theoretical Physics, Chinese Academy of Sciences,
Beijing 100190, P. R. China}

\author{Tianjun Li}

\affiliation{State Key Laboratory of Theoretical Physics,
 Institute of Theoretical Physics, Chinese Academy of Sciences,
Beijing 100190, P. R. China}

\affiliation{School of Physical Electronics,
University of Electronic Science and Technology of China,
Chengdu 610054, P. R. China}

\author{Xiao-Chuan Wang}

\affiliation{State Key Laboratory of Theoretical Physics,
 Institute of Theoretical Physics, Chinese Academy of Sciences,
Beijing 100190, P. R. China}

\author{Xiao-Hong Wu}

\affiliation{Institute of Modern Physics, East China University of Science
and Technology, 130 Meilong Road, Shanghai 200237, P. R. China}

\affiliation{State Key Laboratory of Theoretical Physics,
 Institute of Theoretical Physics, Chinese Academy of Sciences,
Beijing 100190, P. R. China}

\date{September 9, 2014}

\begin{abstract}

We propose a class of the two Higgs doublet Standard models (SMs) with 
a SM singlet and a class of supersymmetric SMs with two pairs of Higgs 
doublets, where the right-handed up/charm quarks and the right-handed 
top quark have different quantum numbers under extra discrete symmetries. 
Thus, the right-handed up and charm quarks couple 
to one Higgs doublet field, while the right-handed top quark couples 
to another Higgs doublet. 
The quark CKM mixings can be generated from 
the down-type quark sector. As one of phenomenological consequences 
in our models, we explore whether one can accommodate the observed direct 
CP asymmetry difference in singly Cabibbo-suppressed D decays. We show 
that it is possible to explain the measured values of CP violation 
under relevant experimental constraints.

\end{abstract}



\maketitle

\section{Introduction}

Experimental data from the ATLAS~\cite{2012c, 2013c}, CMS~\cite{12c, 13c}, 
D0 and CDF~\cite{Cdf:d0}  Collaborations have confirmed the existence of 
the Standard Model (SM) Higgs boson. However, 
the quark CKM mixing phase is not enough to explain the baryon 
asymmetry in the Universe and gives the contributions to electric dipole moments (EDMs)
of electron and neutron much smaller than the experimental limits.
Therefore, one needs new sources of CP violation, which has been one
of the main motivations to search for new theoretical models beyond the SM for a long time.

The minimal extension of the SM is to enlarge the Higgs sector~\cite{Lee}. It has been shown 
that the two-Higgs-doublet models (2HDMs) naturally 
accommodates the electroweak precision tests, giving rise at the same time
to many interesting phenomenological effects~\cite{gb}.
For a recent review on 
two-Higgs-doublet SMs, please see~\cite{Branco:2011iw}. 
The generic scalar spectrum of the two-Higgs-doublet models consists of
three neutral Higgs bosons and one charged Higgs boson pair.
The direct searches for additional scalar particles at the LHC 
or indirect searches via precision flavor experiments will therefore continue
being an important task in the following years.

In this paper, we will propose a class of the two Higgs doublet SMs with a SM singlet
and a class of the supersymmetric SMs with two pairs of Higgs doublets, where the 
right-handed up/charm quarks and right-handed top quark have different quantum numbers 
under extra discrete symmetries. Therefore, the right-handed up and charm quarks couple to 
one Higgs doublet field, while the right-handed top quark couples to another 
Higgs doublet due to additional discrete symmetries. All the down-type quarks 
couple to the same Higgs doublet, and all the charged leptons couple to 
the same Higgs doublet. Also, the quark CKM mixings can be
generated from the down-type quark sector.
In particular, the first two-generation 
up-type quarks can have relatively large Yukawa couplings.
As one of the phenomenological consequences of our models we explore if 
one can accommodate the experimental measurement of direct CP asymmetry 
difference in singly Cabibbo-suppressed $D$ decays.

The CP asymmetry difference in $D^0 \to K^+ K^-$ and 
$D^0 \to \pi^+ \pi^-$ decays has been measured
by the LHCb Collaboration~\cite{Aaij:2011in}.
Combined with the results from the CDF~\cite{Collaboration:2012qw},
Belle~\cite{Ko:2012px}, and previous BaBar~\cite{Aubert:2007if} Collaborations,
the Heavy Flavor Averaging Group yields a world average of the difference
of direct CP asymmetry in $D^0 \to K^+ K^-$ and $D^0 \to \pi^+ \pi^-$ decays,
$\Delta A_{CP} = (-0.656 \pm 0.154)\%$ in March 2012~\cite{Amhis:2012bh}.
However, the above results have not been confirmed by the latest experimental measurements.
The updated LHCb result with pion-tagged analysis gives
$\Delta A_{CP} = (-0.34 \pm 0.15 \pm 0.10)\%$~\cite{LHCb2013pion}.
For the muon tagging, the measurements from LHCb
using $1.0 fb^{-1}$ data at 7 TeV have
$\Delta A_{CP} = (0.4\pm 0.3\pm 0.14)\%$~\cite{Aaij:2013bra},
and $\Delta A_{CP} = (+ 0.14\pm 0.16\pm 0.08)\%$~\cite{Aaij:2014gsa}
with the latest $3 fb^{-1}$ data,
which have an opposite sign compared to the pion-tagged results.
In combination, the current world-averaged direct charm meson CP violation
is $\Delta A_{CP} = (0.253 \pm 0.104)\%$
from the Heavy Flavor Averaging Group~\cite{Amhis:2012bh}.

The CP asymmetry in charm meson decays has inspired
a lot of theoretical discussions. The SM contributions to the direct CP asymmetry
are discussed in Refs.~\cite{Isidori:2011qw,Li:2012cfa,Cheng:2012wr}. Li et al~\cite{Li:2012cfa} 
showed that $\Delta A_{CP} = A_{CP}(K^+ K^-) -A_CP(\pi^+ \pi^- )= -1.00 \times 10^{-3} $, which is lower than 
the LHCb and CDF data. Based on the topological diagram approach for tree-level amplitudes 
and QCD factorization for a crude estimation of perturbative penguin amplitudes, 
Cheng and Chiang~\cite{Cheng:2012wr} showed that the CP asymmetry difference $
\Delta A_{CP}$ is of order $-(0.14 \sim 0.15)\%.$  Even with the maximal magnitude 
of QCD-penguin exchange amplitude $|PE| \sim T$ ($T$ is the tree-level amplitude) and 
a maximal strong phase relative to $T$, one can only get
$\Delta A_{CP} = -0.25 \%$ which is still lower than the current world average.
The $SU(3)$ effects have also been studied~\cite{Pirtskhalava:2011va,Bhattacharya:2012ah,
Franco:2012ck,Cheng:2012xb,Hiller:2012xm}. For the recent discussions 
on the subjects, please see Ref.~\cite{Brod:2013cua}. While the experiment is still not conclusive, 
there are some attempts to estimate the effects from new physics models,
e.g., fourth generation~\cite{Rozanov:2011gj}, left-right model~\cite{Chen:2012usa},
diquark~\cite{Chen:2012am}, supersymmetry~\cite{Chang:2012gna,Giudice:2012qq},
Randall-Sundrum model~\cite{Delaunay:2012cz}, compositeness~\cite{KerenZur:2012fr,DaRold:2012sz},
minimal flavor violation~\cite{bbhl:2013b},
other new physics models~\cite{Altmannshofer:2012ur}, and a $\chi^2$ analysis of different measurements
in the charm system~\cite{Dighe:2013epa}.

We calculate the direct CP asymmetry difference in charm meson decays with experimental 
constraints satisfied in our models in the paper. The new feature of our work is that
we consider the contributions from Higgs penguin induced operators,
and the mixing effect of Higgs penguin induced operator $O_{13}$
into chromomagnetic operator $O_{8g}$ at charm mass $m_c$ scale. We find 
that it is possible to explain the measured values of CP violation 
under relevant experimental constraints.


This paper is organized as follows. We present a class
of two-Higgs-doublet SMs and a class of the supersymmetric SMs in
Sections II and III. The effective Lagrangian of $c \to u$ transition,
relevant Wilson coefficients, direct CP asymmetry in charm meson decays,
and $\Delta c=2$ and $\Delta c=1$ constraints are given in Section IV.
We conclude in Section V.

\section{Nonsupersymmetric SMs}

We consider the two-Higgs-doublet Standard Models~\cite{Lee}.
First, let us explain the convention. We denote the
left-handed quark doublets, the right-handed up-type quarks,
the right-handed down-type quarks, the left-handed lepton doublets,
and the right-handed leptons as $q_i$, $u_i$, $d_i$, $l_i$, and $e_i$,
respectively, where $i=1, ~2, ~3$. In addition, we introduce
two pairs of the Higgs doublets as $\phi_1$ and $\phi_2$,
and a SM singlet Higgs field $S$. Following the common convention,
we assume that the $U(1)_Y$ charges for both $\phi_1$ and
$\phi_2$ are $+1$.

Without loss of generality, we assume that $\phi_1$ couples
to the right-handed up and charm quarks, while $\phi_2$ couples
to the right-handed top quark. We classify the models as follows
\begin{itemize}

\item {Model I: both the down-type quarks and the charged leptons couple
to $\phi_2$.}

\item {Model II: the down-type quarks couple to $\phi_1$ while
the charged letpons couple to $\phi_2$.}

\item {Model III: the charged letpons couple to $\phi_1$ while
the down-type quarks couple to $\phi_2$.}

\item {Model IV: both the down-type quarks and charged leptons couple
to $\phi_1$.}

\end{itemize}

To avoid the flavour changing neutral current (FCNC) constraints~\cite{Glashow:1976nt}, 
we introduce a $Z_3$ symmetry. Under this $Z_3$ symmetry, the quark doublets, 
the up-type quarks, the Higgs fields, and the singlet transform as follows
\begin{eqnarray}
q_i  \leftrightarrow q_i~,~u_k  \leftrightarrow u_k~,~t  \leftrightarrow \omega t~,~
\phi_1  \leftrightarrow  \phi_1~,~\phi_2  \leftrightarrow \omega \phi_2~,~
S \leftrightarrow \omega S~,~\,
\end{eqnarray}
where $\omega^3=1$, $i=1,~2,~3$, and $k=1,~2$. The transformation properties for
down-type quarks, lepton doublets, and charged leptons will be given later
for each model. By the way, to escape the FCNC constraints in the nonsupersymmetric SMs, 
we just need to consider $Z_2$ symmetry, {\it i.e.}, we change each ``$\omega^2$'' and ``$\omega$'' 
into the ``$-$'' sign in our transformation equations. To match the supersymmetric SMs, we consider
the $Z_3$ symmetry in this paper.

\subsection{Model I}

Under this $Z_3$ symmetry, the down-type quarks, the lepton doublets, and the charged leptons
transform as follows
\begin{eqnarray}
d_i  \leftrightarrow  \omega^2 d_i~,~l_i  \leftrightarrow l_i~,~
e_i  \leftrightarrow \omega^2 e_i~.~\,
\end{eqnarray}
Then, the SM fermion Yukawa Lagrangian is
\begin{eqnarray}
-{\cal L} &=& y_{ki}^u\overline{u}_k q_i \phi_1 + y_{i}^t\overline{t} q_i \phi_2
+ y_{ij}^d \overline{d}_i q_j
\tilde{\phi}_2 + y_{ij}^e \overline{e}_i l_j \tilde{\phi}_2 + {\rm H.C.}~,~\,
\end{eqnarray}
where $y_{ij}^u$, $y_{ij}^d$ and $y_{ij}^e$ are Yukawa couplings,
and $\tilde{\phi}_i = i\sigma_2 \phi_i^*$. Here, $\sigma_2$ is
the second Pauli matrix. In particular, to avoid the
 FCNC constraints~\cite{Glashow:1976nt}, we assume that
the Yukawa couplings $y_{13}^u$, $y_{23}^u$, $ y_{1}^t$ and $ y_{2}^t$
are relatively small. It is clear that in the limit
 $y_{13}^u=y_{23}^u=y_{1}^t=y_{2}^t=0$, there is no FCNC effect.
Moreover, the quark CKM mixings are generated from the down-type quark
sector. Let us define
\begin{eqnarray}
\tan\beta \equiv {{<\phi_2>}\over {<\phi_1>}}~.~\,
\end{eqnarray}
At large $\tan\beta$, the Higgs fields with dominant components
from $\phi_1$ will have large Yukawa couplings with the first 
two-generation up-type quarks.

The most general renormalizable Higgs potential at tree level, which
is invariant under the $SU(2)_L\times U(1)_Y$ gauge symmetry and
the $Z_3$ symmetry, is
\begin{eqnarray}
V &=& {{\lambda_1}\over 2} (\phi^{\dagger}_1 \phi_1)^2
  + {{\lambda_2}\over 2} (\phi^{\dagger}_2 \phi_2)^2
  + {{\lambda_S}\over 2} (S^{\dagger} S)^2
  + {{\lambda_3}\over 2} (\phi^{\dagger}_1 \phi_1)
    (\phi^{\dagger}_2 \phi_2)
  + {{\lambda_4}\over 2} (\phi^{\dagger}_1 \phi_2)
    (\phi^{\dagger}_2 \phi_1)
\nonumber \\  &&
  + {{\lambda_{S1}}\over 2} (S^{\dagger} S)
    (\phi^{\dagger}_1 \phi_1)
  + {{\lambda_{S2}}\over 2} (S^{\dagger} S)
     (\phi^{\dagger}_2 \phi_2)
+\left[ A S \phi^{\dagger}_2  \phi_1
    +{\rm H.C.} \right]
\nonumber \\  &&
-  {1\over 2} m_{11}^2 \phi^{\dagger}_1 \phi_1
   -  {1\over 2} m_{22}^2 \phi^{\dagger}_2 \phi_2
   -  {1\over 2} m_{S}^2 S^{\dagger} S ~,~\,
\label{NS-Potential}
\end{eqnarray}
where $\lambda_i$, $\lambda_S$, $\lambda_{S1}$, and $\lambda_{S2}$
are dimensionless parameters, $m_{11}^2$, $ m_{22}^2$, and $m_{S}^2$
are mass parameters, and $A$ is a mass dimension-one
parameter which is similar to the supersymmetry breaking
trilinear soft term.
 $\lambda_i$ for $i=1, 2, 3, 4$, $\lambda_S$,
$\lambda_{S1}$, $\lambda_{S2}$, $m_{11}^2$, $m_{22}^2$ and
$ m_{S}^2$ are real, while  $A$ is complex.
In addition, the term
$\lambda_5 (\phi^{\dagger}_1 \phi_2)^2 $ 
and its Hermitian conjugate, are forbidden by discrete
$Z_3$ symmetry. Also, the terms
$\lambda_6 (\phi^{\dagger}_1 \phi_1) (\phi^{\dagger}_1 \phi_2) $ and
 $\lambda'_6 (\phi^{\dagger}_2 \phi_2) (\phi^{\dagger}_1 \phi_2) $,
as well as their Hermitian conjugates, which will induce the
FCNC processes~\cite{Glashow:1976nt}, are forbidden in our model, too.
Interestingly, our model can be consistent with the constraints
from the CP violation and FCNC processes even if  $A$ is  not
real~\cite{Akeroyd:2000wc, Ginzburg:2003fe, Horejsi:2005da, Ginzburg:2005dt}.

For simplicity, we assume that the up-type quark Yukawa matrix is diagonal,
and then there are no tree-level FCNC processes. Also, we assume that
$A$ is relatively small, and the vacuum expectation value (VEV) 
of $S$ is much larger than the VEVs
of $\phi_1$ and $\phi_2$, for example, $\langle S \rangle \simeq 3$~TeV.  
Thus, the mixings between $S$ and $\phi_i$ are small and can be neglected.
The Lagrangian of relevance for our discussion of direct CP violation
in charm meson decays can be written as
\begin{eqnarray*}
-{\cal L} &=&
\frac{g m_{u_k}}{2m_W}\frac{c_\alpha}{c_\beta}H\overline{u}_k u_k-
 \frac{g m_{u_k}}{2m_W}\frac{s_\alpha}{c_\beta}h\overline{u}_k u_k
+\frac{g m_t}{2m_W}\frac{s_\alpha}{s_\beta}H\overline{t} t
+\frac{g m_t}{2m_W}\frac{c_\alpha}{s_\beta}h\overline{t} t \\
&&-\frac{g m_{d_j}}{2m_W}\frac{s_\alpha}{s_\beta}H\overline{d}_j d_j
-\frac{g m_{d_j}}{2m_W}\frac{c_\alpha}{s_\beta}h\overline{d}_j d_j \\
&&+i \frac{g m_{u_k}}{2m_W} t_\beta A\overline{u}_k \gamma^{5} u_k
+i \frac{g m_t}{2m_W} ct_\beta A\overline{t} \gamma^{5} t
+i \frac{gm_{d_j}}{2m_W}ct_\beta A\overline{d}_j \gamma^{5} d_j \\
&&+\frac{g m_{u_k}}{2m_W} V_{kj} t_\beta H^+ \overline{u}_k P_L d_j
-\frac{gm_{d_j}}{2m_W} V_{kj} ct_\beta H^+ \overline{u}_k P_R d_j \\
&& -\frac{g m_t}{2m_W} V_{3j} ct_\beta H^+ \overline{t} P_L d_j
-\frac{g m_{d_j}}{2m_W} V_{3j} ct_\beta H^+ \overline{t} P_R d_j + ... ~,~\,
\end{eqnarray*}
where $s_\alpha=\sin\alpha$, $c_\alpha=\cos\alpha$,
$s_\beta=\sin\beta$, $c_\beta=\cos\beta$,
$t_\beta=\tan\beta$, and $ct_\beta=\cot\beta$, with $\alpha$ being the mixing
angle between the real components of $\phi_1^0$ and $\phi_2^0$.

\subsection{Model II}

 Under this $Z_3$ symmetry, the  down-type quarks, lepton doublets, and charged leptons
 transform as follows
\begin{eqnarray}
d_i  \leftrightarrow   d_i~,~l_i  \leftrightarrow l_i~,~e_i  \leftrightarrow \omega^2 e_i~.~\,
\end{eqnarray}
So the SM fermion Yukawa Lagrangian is
\begin{eqnarray}
-{\cal L} &=& y_{ki}^u\overline{u}_k q_i \phi_1 + y_{i}^t\overline{t} q_i \phi_2
+ y_{ij}^d \overline{d}_i q_j
\tilde{\phi}_1 + y_{ij}^e \overline{e}_i l_j \tilde{\phi}_2 + {\rm H.C.}~.~\,
\end{eqnarray}
Similar to Model I, we assume that
the Yukawa couplings $y_{13}^u$, $y_{23}^u$, $ y_{1}^t$ and $ y_{2}^t$
are relatively small. The most general renormalizable Higgs potential 
at tree level, which is invariant under the $SU(2)_L\times U(1)_Y$ gauge symmetry 
and the $Z_3$ symmetry, is the same as that in Eq.~(\ref{NS-Potential}) in Model I.
At large $\tan\beta$, the Higgs fields with dominant components
from $\phi_1$ will have large Yukawa couplings with the first 
two-generation up-type quarks, and all the down-type quarks.

With the same assumptions as in Model I,
the Lagrangian of relevance for our discussion can be written as
\begin{eqnarray*}
-{\cal L} &=&
\frac{g m_{u_k}}{2m_W}\frac{c_\alpha}{c_\beta}H\overline{u}_k u_k-
 \frac{g m_{u_k}}{2m_W}\frac{s_\alpha}{c_\beta}h\overline{u}_k u_k
+\frac{g m_t}{2m_W}\frac{s_\alpha}{s_\beta}H\overline{t} t
+\frac{g m_t}{2m_W}\frac{c_\alpha}{s_\beta}h\overline{t} t \\
&&-\frac{g m_{d_j}}{2m_W}\frac{c_\alpha}{c_\beta}H\overline{d}_j d_j
+\frac{g m_{d_j}}{2m_W}\frac{s_\alpha}{c_\beta}h\overline{d}_j d_j \\
&&+i \frac{g m_{u_k}}{2m_W} t_\beta A\overline{u}_k \gamma^{5} u_k
+i \frac{g m_t}{2m_W} ct_\beta A\overline{t} \gamma^{5} t
-i \frac{g m_{d_j}}{2m_W} t_\beta A\overline{d}_j \gamma^{5} d_j \\
&&+ \frac{g m_{u_k}}{2m_W} V_{kj} t_\beta H^+ \overline{u}_k P_L d_j
+ \frac{gm_{d_j}}{2m_W} V_{kj} t_\beta H^+ \overline{u}_k P_R d_j \\
&& - \frac{g m_t}{2m_W} V_{3j} ct_\beta H^+ \overline{t} P_L d_j
+ \frac{g m_{d_j}}{2m_W} V_{3j} t_\beta H^+ \overline{t} P_R d_j + ... ~.~\,
\end{eqnarray*}

\subsection{Model III}

Under this $Z_3$ symmetry, the down-type quarks, the lepton doublets, and the charged leptons
 transform as follows
\begin{eqnarray}
d_i  \leftrightarrow \omega^2 d_i~,~l_i  \leftrightarrow l_i~,~
e_i  \leftrightarrow  e_i~.~\,
\end{eqnarray}
So the SM fermion Yukawa Lagrangian is
\begin{eqnarray}
-{\cal L} &=& y_{ki}^u\overline{u}_k q_i \phi_1 + y_{i}^t\overline{t} q_i \phi_2
+ y_{ij}^d \overline{d}_i q_j
\tilde{\phi}_2 + y_{ij}^e \overline{e}_i l_j \tilde{\phi}_1 + {\rm H.C.}~.~\,
\end{eqnarray}
At large $\tan\beta$, the Higgs fields with dominant components
from $\phi_1$ will have large Yukawa couplings
with the first two-generation up-type quarks, and all the charged leptons.
The rest discussion is similar to those in Models I and II.

\subsection{Model IV}

Under this $Z_3$ symmetry, the down-type quarks, the lepton doublets, and the charged leptons
 transform as follows
\begin{eqnarray}
d_i  \leftrightarrow   d_i~,~l_i  \leftrightarrow l_i~,~
e_i  \leftrightarrow  e_i~.~\,
\end{eqnarray}
Then, the SM fermion Yukawa Lagrangian is
\begin{eqnarray}
-{\cal L} &=& y_{ki}^u\overline{u}_k q_i \phi_1 + y_{i}^t\overline{t} q_i \phi_2
+ y_{ij}^d \overline{d}_i q_j
\tilde{\phi}_1 + y_{ij}^e \overline{e}_i l_j \tilde{\phi}_1 + {\rm H.C.}~.~\,
\end{eqnarray}
At large $\tan\beta$, the Higgs fields with dominant components
from $\phi_1$ will have large Yukawa couplings
with the first two-generation up-type quarks, all the down-type quarks,
and all the charged leptons.
The rest discussion is similar to those in Models I and II.

\section{Supersymmetric Standard Models}

First, let us explain the convention. We denote the chiral superfields
for the quark doublets, the right-handed up-type quarks,
the right-handed down-type quarks, the lepton doublets,
and the right-handed charged 
leptons as $Q_i$, $U_i^c$, $D_i^c$, $L_i$, and $E_i^c$,
respectively, where $i=1, ~2, ~3$. We also introduce two pairs of
Higgs doublets ($H_u$, $H_d$), and ($H'_u$, $H'_d$).
In addition, we introduce three SM singlet Higgs fields
$S$, $S'$ and $T$.

Without loss of generality, we assume that $H_u$ couples
to the right-handed up and charm quarks, $H'_u$ couples
to the right-handed top quark, and $H_d$ couples to the
right-handed down-type quarks. We classify the models as follows
\begin{itemize}

\item {Model A: $H'_d$ couples to the charged letpons.}

\item {Model B: $H_d$ couples to the charged letpons.}

\end{itemize}

To solve the $\mu$ problem, we consider a $Z_3\times Z'_3$
discrete symmetry. Under the $Z_3$ symmetry, the SM quarks,
the Higgs fields, and the singlet fields transform as follows
\begin{eqnarray}
&& Q_i  \leftrightarrow \omega Q_i~,~
U^c_k  \leftrightarrow \omega U^c_k~,~
T^c  \leftrightarrow \omega^2 T^c~,~
D^c_i  \leftrightarrow \omega D^c_i~,~
\nonumber \\  &&
H_{u,d}  \leftrightarrow \omega H_{u,d}~,~
H'_{u,d}  \leftrightarrow  H'_{u,d}~,~
S \leftrightarrow  \omega S~,~\,
S' \leftrightarrow  S'~,~\,
T \leftrightarrow  \omega^2 T~,~\,
\end{eqnarray}
where  $\omega^3=1$.
And under the $Z'_3$ symmetry, the SM quarks,
the Higgs fields, and the singlet fields transform as below
\begin{eqnarray}
&& Q_i  \leftrightarrow  Q_i~,~
U^c_i  \leftrightarrow  U^c_i~,~
T^c  \leftrightarrow \omega^{\prime 2} T^c~,~
D^c_i  \leftrightarrow  D^c_i~,~
\nonumber \\  &&
H_{u,d}  \leftrightarrow  H_{u,d}~,~
H'_{u,d}  \leftrightarrow \omega' H'_{u,d}~,~
S \leftrightarrow  S~,~\,
S' \leftrightarrow  \omega'  S'~,~\,
T \leftrightarrow  \omega^{\prime 2} T~,~\,\,
\end{eqnarray}
where $\omega^{\prime 3}=1$.

\subsection{Model A}

 Under the $Z_3\times Z'_3$ symmetry, the lepton doublets
and the charged leptons, respectively, transform as follows
\begin{eqnarray}
&& L_i  \leftrightarrow   L_i~,~
E^c_i  \leftrightarrow  E^c_i~,~
\nonumber \\  &&
L_i  \leftrightarrow   \omega' L_i~,~
E^c_i  \leftrightarrow  \omega' E^c_i~.~
\end{eqnarray}

Then, the SM fermion Yukawa Lagrangian is
\begin{eqnarray}
W_{\rm Yukawa} &=& y_{ik}^u Q_i H_u U^c_k + y^t_i Q_i T^c H'_u+
 y_{ij}^d Q_i H_d D^c_j + y_{ij}^e L_i H'_d E^c_j
\nonumber \\  &&
+ \lambda_1 S H_d H_u
+ \lambda_2 S' H'_d H_u'
+ \lambda_3 T H_d H_u' +
\lambda_4 T H_d' H_u
\nonumber \\  &&
+ \lambda_5 S S' T
+ {{\kappa_{1}}\over 3} S^3 + {{\kappa_{2}}\over 3} S^{\prime 3}
+ {{\kappa_{3}}\over 3} T^3
~,~\,
\end{eqnarray}
where $y_{ik}^u$, $y^t_i$,
 $y_{ij}^d$, $y_{ij}^e$, $\lambda_i$,
and $\kappa_i$ are Yukawa couplings.
To avoid the FCNC constraints, we assume that the
Yukawa couplings $y_{31}^u$, $y_{32}^u$, $y^t_1$
and  $y^t_2$ are relatively small, similar to the
nonsupersymmetric models. 
In our model, we define
\begin{eqnarray}
\tan\beta \equiv {{<H_d>}\over {<H_u>}}~,~\,
\end{eqnarray}
which is different from the traditional minimal supersymmetric standard model.
The VEV of $H_u$ can be much smaller than that of $H_d$,
since  $H_u^\prime$ couples to the top quark, {\it i.e.}, 
the charm Yukawa coupling can be order $1$.
Note that the VEV of $H_d$ can be about one order
larger that that of $H'_d$, and we obtain that
the Yukawa couplings of down-type quarks can
be about one order smaller than those of charged leptons
compared to the SM.

\subsection{Model B}

Under the $Z_3\times Z'_3$ symmetry, the lepton doublets
and the charged leptons, respectively, transform as follows
\begin{eqnarray}
&& L_i  \leftrightarrow  \omega L_i~,~
E^c_i  \leftrightarrow \omega E^c_i~,~
\nonumber \\  &&
L_i  \leftrightarrow    L_i~,~
E^c_i  \leftrightarrow   E^c_i~.~
\end{eqnarray}

Then, the SM fermion Yukawa Lagrangian is
\begin{eqnarray}
W_{\rm Yukawa} &=& y_{ik}^u Q_i H_u U^c_k + y^t_i Q_i T^c H'_u+
 y_{ij}^d Q_i H_d D^c_j + y_{ij}^e L_i H_d E^c_j
\nonumber \\  &&
+ \lambda_1 S H_d H_u
+ \lambda_2 S' H'_d H_u'
+ \lambda_3 T H_d H_u' +
\lambda_4 T H_d' H_u
\nonumber \\  &&
+ \lambda_5 S S' T
+ {{\kappa_{1}}\over 3} S^3 + {{\kappa_{2}}\over 3} S^{\prime 3}
+ {{\kappa_{3}}\over 3} T^3
~.~\,
\end{eqnarray}
To avoid the FCNC constraints, similar to Model A, we assume that the
Yukawa couplings $y_{31}^u$, $y_{32}^u$, $y^t_1$
and  $y^t_2$ are relatively small.

\section{Effective Hamiltonian and Direct CP Asymmetries in $D$ Meson Decays}

The effective Hamiltonian for the $c \to u$ transition can be written as
\begin{eqnarray}
\mathcal{H}_{\Delta C=1}^{eff} &=& \frac{G_F}{\sqrt{2}}
 \biggl\{\sum_{p=d,s} \lambda_p (C_1^p O_1^p + C_2^p O_2^p) \nonumber \\
&+&\lambda_b \left[\sum_{i=3}^6 C_i O_i + C_{7\gamma} O_{7\gamma}
  + C_{8g} O_{8g}
 + \sum_{i=11}^{16} \sum_{q=u,d,s,c} C_i^q O_i^q \right] \biggr\} ,
\end{eqnarray}
with $\lambda_{p}=V_{cp}^{\ast} V_{up}$ ($p=d,s$)
 and $\lambda_{b}=V_{cb}^{\ast} V_{ub}$.

The complete list of operators is given as follows
\begin{eqnarray}
O_1^p &=& (\bar{u} p)_{V-A}(\bar{p} c)_{V-A}, \nonumber\\
O_2^p &=& (\bar{u}_\alpha p_\beta)_{V-A}(\bar{p}_\beta c_\alpha)_{V-A},\nonumber\\
O_3 &=& (\bar{u} c)_{V-A} \sum_q (\bar{q} q)_{V-A}, \nonumber\\
O_4 &=& (\bar{u}_\alpha c_\beta)_{V-A} \sum_q (\bar{q}_\beta q_\alpha)_{V-A},\nonumber\\
O_5 &=& (\bar{u} c)_{V-A} \sum_q (\bar{q} q)_{V+A}, \nonumber\\
O_6 &=& (\bar{u}_\alpha c_\beta)_{V-A} \sum_{q}(\bar{q}_\beta q_\alpha)_{V+A},\nonumber\\
O_{7\gamma}  &=& \frac{e}{8\pi^2} m_c [\bar{u} \sigma_{\mu\nu} (1+\gamma^5) c] F^{\mu\nu}, \nonumber\\
O_{8g} &=& \frac{g_s}{8\pi^2} m_c [\bar{u} \sigma_{\mu\nu} T^a (1+\gamma^5) c] G_a^{\mu\nu}, \nonumber\\
O_{11}^q &=& (\bar{u} c)_{S+P} (\bar{q} q)_{S-P}, \nonumber\\
O_{12}^q &=& (\bar{u}_\alpha c_\beta)_{S+P} (\bar{q}_\beta q_\alpha)_{S-P}, \nonumber\\
O_{13}^q &=& (\bar{u} c)_{S+P} (\bar{q} q)_{S+P}, \nonumber\\
O_{14}^q &=& (\bar{u}_\alpha c_\beta)_{S+P} (\bar{q}_\beta q_\alpha)_{S+P}, \nonumber\\
O_{15}^q &=& [\bar{u} \sigma_{\mu\nu} (1+\gamma^5)c] [\bar{q}\sigma^{\mu\nu}(1+\gamma^5) q], \nonumber\\
O_{16}^q &=& [\bar{u}_\alpha \sigma_{\mu\nu}(1+\gamma^5)c_\beta] [\bar{q}_\beta\sigma^{\mu\nu}(1+\gamma^5) q_\alpha] ,
\end{eqnarray}
with $V \pm A = \gamma^\mu (1 \pm \gamma^5)$ and $S \pm P = (1 \pm \gamma^5)$.

The direct CP asymmetry of $D^0 \to K^+ K^-$ can be written as
\begin{eqnarray}
a_{K^+ K^-} &=&
  2 {\rm Im} \bigg(\frac{\lambda_b}{\lambda_s} R^s_{K,\rm SM}\bigg) +
  2 {\rm Im} \bigg(\frac{\lambda_b}{\lambda_s} R^s_{K,\rm NP}\bigg),
\end{eqnarray}
where
\begin{eqnarray}
R^s_{K,\rm SM} &=& \frac{a_4^{\rm SM}+r_\chi a_6^{\rm SM}}{a_1},
R^s_{K,\rm NP} = \frac{1}{a_1} \bigg( a_4^{\rm NP} - \frac{1}{12} a_{12}^s +
   r_\chi (a_6^{\rm NP} + \frac{1}{4} a_{14}^s + 3 a_{16}^s ) \bigg),
\end{eqnarray}
where maximal strong phase is assumed, and only weak phase is included
in the above equation.
The $a_i$ coefficients are estimated in naive factorization
\begin{eqnarray}
a_4^{\rm NP} &=& 3 a_6^{\rm NP} =
 - \frac{3 C_F \alpha_s}{2 \pi N_C} C_{8g}^{\rm NP} ~,~\, \nonumber\\
a_{12}^s &=& C_{12}^s + C_{11}^s/N_C ~,~\,\nonumber\\
a_{14}^s &=& C_{14}^s + C_{13}^s/N_C ~,~\,\nonumber\\
a_{16}^s &=& C_{16}^s + C_{15}^s/N_C ~,
\end{eqnarray}
where the Wilson coefficients $C_{8g,11,12,13,14,15,16}$
are evaluated at charm quark mass $m_c$ scale.
For the direct CP asymmetry of $D^0 \to \pi^+ \pi^-$,
the upper index $s$ should be replaced with $d$.
In the flavor $SU(3)$ limit, we have $a_{\pi^+\pi^-} \simeq - a_{K^+K^-}$.

The Wilson coefficients can be evolved from $W$ boson mass $m_w$ scale to $m_c$ scale
through the intermediate bottom quark mass scale $m_b$~\cite{Buchalla:1995vs}.
The main contribution in our case
is $C_{8g}(m_c)$, which can be written as~\cite{Borzumati:1999qt,
Hiller:2003js,Cheng:2003im,Cheng:2004jf}
\begin{eqnarray}
C_{8g}(m_c) &\simeq& 0.4983 C_{8g}(m_w) - 0.1382 C_2(m_w)
  + 0.4922 C_{13}^c(m_w).
\end{eqnarray}

The direct CP asymmetry in the decays $D^0 \to K^+ K^-$
and $D^0 \to \pi^+ \pi^-$ can be estimated as
\begin{eqnarray}
\Delta a_{CP} &=& a_{K^+K^-} - a_{\pi^+\pi^-} \nonumber\\
  & \simeq & [-0.01676 C_{8g}^{\rm NP}(m_w) + 0.1142 C_{13}(m_w) ] \times 1\% .
\end{eqnarray}
For $\Delta a_{CP} \sim 0.1\%$, we should have $C_{8g}^{\rm NP}(m_w) \sim 10$,
or $C_{13}(m_w) \sim 1$.

We can further express $C_{11,13}^c$ as~\cite{Cheng:2004jf}
\begin{eqnarray}
C_{11}^c &=& \frac{e^2}{16\pi^2} ( C_{Q_1}^c - C_{Q_2}^c ) ~,~~~
C_{13}^c =   \frac{e^2}{16\pi^2} ( C_{Q_1}^c + C_{Q_2}^c ) .
\end{eqnarray}

To follow, we will calculate $C_{Q_{1,2}}^c$ and $C_{8g}$
at $m_w$ scale in Models I, II, and A.

The contributions to $C_{8g}$ from charged Higgs boson exchanges are
\begin{eqnarray}
C_{8g} &=& - \cot^2_\beta \frac{1}{6} D(x_{H^\pm}) - E(x_{H^\pm}) 
\end{eqnarray}
in Model I, and
\begin{eqnarray}
C_{8g} &=& t^2_\beta \bigg[- \frac{1}{6} D(x_{H^\pm}) - E(x_{H^\pm}) \bigg] 
\end{eqnarray}
in Model II, with $x_{H^\pm} = m_b^2/m_{H^\pm}^2$.
The one-loop functions $D$ and $E$ are defined in Ref.~\cite{Grinstein:1990tj}.

In our calculations,
we work in the limit of vanishing light quark masses,
$m_u = m_d = m_s = 0$.
The Wilson coefficients $C_{Q_{1,2}}$ at the leading order of
${\cal O}(\tan^2\beta)$ in Model I are
\begin{eqnarray}
C^c_{Q_1} &=& - \frac{m_b^2 m_c^2}{4 m_w^2 s_w^2 c_\beta^2} \bigg(
  \frac{c_\alpha^2}{m_H^2} + \frac{s_\alpha^2}{m_h^2} \bigg)
  \bigg[ f_{b0}(x_{H^\pm}) - f_{b0}(x_W) \bigg] \nonumber\\
  &-& \frac{3}{8} \frac{m_c^2 t_\beta}{s_w^2 c_\beta} \bigg(
  \frac{c_\alpha}{m_H^2}s_{\beta-\alpha}
   + \frac{s_\alpha}{m_h^2} c_{\beta-\alpha} \bigg) f_{c00}(x_W, x_{H^\pm})
 \nonumber\\
  &+& \frac{m_c^2 t_\beta^2}{12 m_{H^\pm}^2 s_w^2} |V_{cb}|^2
   f_{d00}(x_W, x_W, x_{H^\pm}) ~,~\nonumber\\
C^c_{Q_2} &=& \frac{m_b^2 m_c^2 t_\beta^2}{4 m_w^2 s_w^2 m_A^2} \bigg[
   f_{b0}(x_{H^\pm}) - f_{b0}(x_W) \bigg] \nonumber\\
  &+& \frac{3}{8} \frac{m_c^2 t_\beta^2}{s_w^2 m_A^2}
    f_{c00}(x_W, x_{H^\pm}) \nonumber\\
  &-& \frac{m_c^2 t_\beta^2}{12 m_{H^\pm}^2 s_w^2} |V_{cb}|^2
   f_{d00}(x_W, x_W, x_{H^\pm}) ~.~
\end{eqnarray}
where the one-loop functions $f_{b0,c00,d00}$ are defined in Ref.~\cite{Huang:2002ni}.

The Wilson coefficients $C_{Q_{1,2}}$ at the leading order of
${\cal O}(\tan^4\beta)$ in Model II are
\begin{eqnarray}
C^c_{Q_1} &=& - \frac{m_b^2 m_c^2}{4 m_w^2 s_w^2} \frac{t_\beta^2}{c_\beta^2}
  \bigg( \frac{c_\alpha^2}{m_H^2} + \frac{s_\alpha^2}{m_h^2} \bigg) f_{b0}(x_{H^\pm}) \nonumber\\
  &-& \frac{m_b^2 m_c^2}{8 m_w^2 s_w^2} \frac{t_\beta^2}{c_\beta^2}
   \bigg( \frac{c_\alpha^2}{m_H^2} + \frac{s_\alpha^2}{m_h^2} \bigg)
   \bigg[ 3 f_{c00}(x_{H^\pm}) + \frac{m_b^2}{m_{H^\pm}^2} f_{c0}(x_{H^\pm}) \bigg] \nonumber\\
  &+& \frac{m_b^4 m_c^2}{12 m_w^2 s_w^2 m_{H^\pm}^4} t_\beta^4 |V_{cb}|^2
   f_{d0}(x_{H^\pm}) ~,~ \nonumber\\
C^c_{Q_2} &=& \frac{m_b^2 m_c^2}{4 m_w^2 s_w^2} \frac{t_\beta^4}{m_A^2} f_{b0}(x_{H^\pm}) \nonumber\\
   &+& \frac{m_b^2 m_c^2}{8 m_w^2 s_w^2} \frac{t_\beta^4}{m_A^2}
     \bigg[ 3 f_{c00}(x_{H^\pm}) + \frac{m_b^2}{m_{H^\pm}^2} f_{c0}(x_{H^\pm}) \bigg]~.~\,
\end{eqnarray}

The leading contributions to the Wilson coefficients
$C_{8g}$ at the order of ${\cal O}(\tan^0\beta)$,
and $C_{Q_{1,2}}$ at the order of ${\cal O}(\tan^2\beta)$ in Model A
from gluino exchanges are
\begin{eqnarray}
C_{8g} &=& - \frac{1}{72 \lambda_b} \frac{g_s^2}{g^2}
   \frac{m_W^2}{m_{\tilde g}^2} \bigg[
   F_{12}(x_{\tilde g}) \delta^{LL}_{12} +
    F^\prime_{12}(x_{\tilde g}) \delta^{LR}_{12} \delta^{LR\ast}_{22}
   - 8 \frac{m_{\tilde g}}{m_c} F_{34}(x_{\tilde g}) \delta^{LR}_{12}
   - 8 \frac{m_{\tilde g}}{m_c} F^\prime_{34}(x_{\tilde g})
    \delta^{LL}_{12} \delta^{LR}_{22} \bigg] , \nonumber\\
C^c_{Q_1} &=& \frac{4}{3 \lambda_b} \frac{g_s^2}{g^2 s_w^2}
   \frac{m_c m_{\tilde g}}{c_\beta^2} \bigg( \frac{c_\alpha^2}{m_H^2} +
    \frac{s_\alpha^2}{m_h^2} \bigg) f^\prime_b(x_{\tilde g})
     \delta^{LL}_{12} \delta^{LR}_{22} , \nonumber\\
C^c_{Q_2} &=& - \frac{4}{3 \lambda_b} \frac{g_s^2}{g^2 s_w^2}
    \frac{m_c m_{\tilde g}}{m_A^2} t_\beta^2 f^\prime_b(x_{\tilde g})
     \delta^{LL}_{12} \delta^{LR}_{22} ,
\end{eqnarray}
where the one-loop functions are defined in Ref.~\cite{Cheng:2004jf}.

The Higgs sector is subject to strong constraints from
both the Higgs coupling measurements~\cite{Beringer:1900zz},
and the direct heavier Higgs searches at LHC,
in particular, $pp \to \Phi \to \tau^+\tau^-$~\cite{Aad:2012cfr,CMS:gya},
$pp \to \Phi \to \mu^+\mu^-$~\cite{CMS:2012lza}
and $pp \to b\Phi \to bbb$~\cite{Chatrchyan:2013qga} channels,
with $\Phi$ as the neutral Higgs boson.
The implications of the Higgs coupling measurements are studied
in Refs.~\cite{Craig:2013hca} and \cite{Dumont:2014wha} 
with direct heavier Higgs searches
within the 2HDMs.
Besides the up and charm quark Yukawa couplings,
the other Higgs couplings in Model I are the same as in 2HDM 1,
and in Model II are the same as in 2HDM 4~\cite{Craig:2013hca}.
We note that the constraints in the $\beta$ and $\cos(\beta-\alpha)$ plane
are much looser in Model I than those in Model II,
while the latter are tightly around
the alignment limit $\alpha=\beta-\pi/2$.
In the numerical calculations, we consider the large $\tan\beta$ case.
The direct heavier Higgs production channels
through $\tau\tau$ and $\mu\mu$ are suppressed by $\sin^2\alpha$
from Yukawa couplings in both Models I and II,
while the $bb$ channel is suppressed by $\sin^2\alpha$ in Model I,
and enhanced by $\tan^2\beta$ in Model II.

For numerical estimations,
we choose the following parameters in the Higgs sector for Model I:
$t_\beta=50$, $s_\alpha=-0.1$, $m_h = 126$~GeV, $m_H = 180$~GeV,
$m_A=220$~GeV, and $m_{H^\pm}=250$~GeV.
In Model II, the measurement of Br$(B \to X_s \gamma)$
puts a stringent bound on the lower limit of the mass of the charged Higgs,
$m_{H^\pm} \ge 380$~GeV at $95\%$ C.L.~\cite{Hermann:2012fc}.
With a heavy charged Higgs pair, the Higgs sector quickly approaches 
the decoupling limits.
For numerical studies, we choose the following parameters for Model II:
$t_\beta=10$, $s_\alpha=-0.1$, $m_h = 126$GeV,
$m_H \simeq m_A \simeq m_{H^\pm}=380$GeV.
In the supersymmetric version Model A,
the Yukawa couplings are similiar to those in Model I.
We also take the supersymmetric scale
$m_{\tilde g}=m_{\tilde q}=2$ TeV~\cite{Beringer:1900zz}.

The charged Higgs contributions can be calculated as
$C_{8g}^{H^\pm} \simeq - 0.9 \times 10^{-3}$ in Model I,
and $C_{8g}^{H^\pm} \simeq - 0.047$ in Model II.
The contributions to $C_{13}(m_w)$ are suppressed in both Models I and II,
where we have $C^c_{13}(m_w) \sim -5.2 \times 10^{-7}$ in Model I,
and $C^c_{13}(m_w) \sim - 1.95 \times 10^{-8}$ in Model II.
Therefore, due to the experimental constraints,
the charged Higgs contributions
cannot accommodate the direct CP measurement of charm decays.

In Model A, for double insertion of
$(\delta^{LL}_{12} \delta^{LR}_{22})$, we have
$C_{8g}^{\tilde g} \sim 7.19 \times
  \frac{(\delta^{LL}_{12} \delta^{LR}_{22})}{10^{-3}}$
and $C^c_{13} \sim - 1.0  \times
  \frac{(\delta^{LL}_{12} \delta^{LR}_{22})}{10^{-3}}$
from gluino exchange.
For $(\delta^{LL}_{12} \delta^{LR}_{22})$ at the order of $10^{-3}$,
we can have both $C_{8g}$ at the order of $10$
and $C^c_{13}$ of order $1$,
which are possible to accommodate the direct CP measurement of charm decays.

The constraint from the $D^0-\bar{D}^0$ system can be found in Ref.~\cite{Gedalia:2009kh}. 
The nonvanishing Wilson coefficients $z_i$ $(i=1,2,...5)$ are
\begin{eqnarray}
z_2 &=& \frac{g^4}{64\pi^2} \frac{\Lambda_{\rm NP}^2}{m_W^2} |\lambda_b|^2
   \frac{m_c^2}{m_W^2} x_W^2
    [ I_2(x_W,x_W/x_{H^\pm}) - 2 I_3(x_W,x_W/x_{H^\pm}) ]
\end{eqnarray}
at the leading order of ${\cal O} (t_\beta^0)$ in Model I, and 
\begin{eqnarray}
z_2 &=& \frac{g^4}{64\pi^2} \frac{\Lambda_{\rm NP}^2}{m_W^2}
  |\lambda_b|^2 t_\beta^4
   x_W^2 [ \frac{1}{4} I_1(x_W,x_W/x_{H^\pm})
          + \frac{m_c^2}{m_W^2} I_2(x_W,x_W/x_{H^\pm}) ]
\end{eqnarray}
at the leading order of ${\cal O} (t_\beta^4)$ in Model II.
The loop functions $I_{1,2,3}$ are defined in Ref.~\cite{Barger:1989fj}.
We can calculate $z_2$ for the above parameters,
$z_2 \simeq - 1.8 \times 10^{-18}$ in Model I,
and $z_2 \simeq 7.7 \times 10^{-13} (\frac{t_\beta}{10})^4$ in Model II,
which are below the experimental limits.

In Model A, we obtain the gluino contributions
\begin{eqnarray}
z_1 &=& - \frac{\alpha_s^2}{216} (\delta^{LL}_{12})^2
  [ 66 \tilde{f}_6(m^2_{\tilde{q}}/m^2_{\tilde{g}}) +
   24 f_6(m^2_{\tilde{q}}/m^2_{\tilde{g}}) ] , \nonumber\\
\tilde{z}_2 &=& - \frac{\alpha_s^2}{216} (\delta^{LL}_{12} \delta^{LR}_{22})^2
  204 f(x) ,
\end{eqnarray}
for $\Lambda_{\rm NP} = m_{\tilde{g}} $,
where the functions $f_6$ and $\tilde{f}_6$ are given
in Ref.~\cite{Grossman:2006jg},
and $f$ is defined as follows
\begin{eqnarray}
f(x) &=& \frac{60x^4(5+x)\ln(x) - 197x^5 - 25x^4 + 300x^3 - 100x^2 +
     25x -3}{60(x-1)^7} ~.~ \nonumber
\end{eqnarray}
The leading order contributions from $(\delta^{LL}_{12})^2$ are
included in $z_1$.
In the numerical estimations,
we take $\delta^{LR}_{22} = (m_c A_c - m_c \mu \tan\beta)/m_{\tilde{q}}^2
 \simeq - m_c \mu \tan\beta/m_{\tilde{q}}^2 \simeq - 0.015$
(with $\mu \sim 1.2 m_{\tilde{q}}$,
$m_c \sim 0.5$ GeV when running to $m_{\tilde{q}}$ scale),
and $\delta^{LL}_{12} \simeq 0.067$.
With the parameter for Model A, we have
$z_1 \simeq 3.0 \times 10^{-7}
  (\frac{\delta^{LL}_{12}}{0.067})^2$,
and
$\tilde{z}_2 \simeq -3.2 \times 10^{-10}
  (\frac{\delta^{LL}_{12} \delta^{LR}_{22}}{10^{-3}})^2$,
which are below the limits from the constraints of the $D^0-\bar{D}^0$ system.
However, due to the $SU(2)$ gauge invariance,
the left-left up-type squark matrix is related to the down-type one.
And we have $\delta^{LL}_{12} \simeq 0.067$ for down-type squarks,
which does not satisfy the constraints from kaon system
for the imaginary part Im$(\delta^{LL}_{12}) \le 0.023$
with the supersymmetry scale at $2$ TeV~\cite{Ciuchini:1998ix}.
One way out is to consider the contributions of
chirally opposite operators.
We can get similiar results
if the above $\delta^{LL}_{12}$ is replaced
with $\delta^{RR}_{12} \sim 0.067$,
and $\delta^{LR}_{22}$ with $\delta^{LR\ast}_{22} \sim -0.015$.
In this case, the up-type and down-type right-right squark matrixes
are not related. Hence, the constraints from the kaon system are relaxed.

Recently, LHCb Collaboration has measured the leptonic and semileptonic
decays of the charm meson, the upper limits are:
$B(D^0 \to \mu^+ \mu^-) < 6.2 (7.6) \times 10^{-9}$
at $90\%$ ($95\%$) C.L.~\cite{Aaij:2013cza} and
$B(D^+ \to \pi^+ \mu^+ \mu^-) < 7.3 (8.3) \times 10^{-8}$
at $90\%$ ($95\%$) C.L.~\cite{Aaij:2013sua}.
The experimental bound on radiative charm decay is
$B(D^0 \to \gamma \gamma) < 2.2 \times 10^{-6}$ at $90\%$ C.L.
from the BABAR Collaboration~\cite{Lees:2011qz},
and $B(D^0 \to \gamma \gamma) < 4.7 \times 10^{-6}$ at $90\%$ C.L.
from BESIII~\cite{Muramatsu:2012nh}.

The corresponding Wilson coefficients are
\begin{eqnarray}
C_{7\gamma} &=& G(x_{H^\pm}) + \frac{1}{6} \cot^2\beta A(x_{H^\pm})
\end{eqnarray}
in Model I, 
\begin{eqnarray}
C_{7\gamma} &=& t^2_\beta [ G(x_{H^\pm}) +  \frac{1}{6} A(x_{H^\pm}) ]
\end{eqnarray}
in Model II, and 
\begin{eqnarray}
C_9 &=& - \frac{-1 + 4 s_W^2}{s_W^2} \cot^2\beta \frac{x_W}{2} B(x_{H^\pm})
        + \cot^2\beta x_{H^\pm} F(x_{H^\pm}) ~,~\nonumber\\
C_{10} &=& - \frac{1}{s_W^2} \cot^2\beta \frac{x_W}{2} B(x_{H^\pm})
\end{eqnarray}
in Model I, while replacing $\cot^2\beta$ with $t^2_\beta$ in Model II.
The functions A, B, G, and F for the $c \to u$ transitions are defined as
\begin{eqnarray}
A(x) &=& -\frac{x}{12} \big( \frac{5 - 10 x - 7 x^2}{(1-x)^3}
   + \frac{6 x (1-3x) \ln x}{(1-x)^4}  \big)~, \nonumber\\
B(x) &=& - \frac{x}{4} \big( \frac{1}{1-x} + \frac{\ln x}{(1-x)^2} \big)~,
\nonumber\\
F(x) &=& \frac{11 - 25x + 40 x^2}{54 (1-x)^3}
   + \frac{2 - 3x + 3x^3}{18 (1-x)^4} ~,\nonumber\\
G(x) &=& - \frac{x}{6} \big( \frac{2}{(1-x)^2}
   - \frac{(1-3x) \ln x}{(1-x)^3} \big) ~,
\end{eqnarray}
which differ from the ones in Ref.~\cite{Grinstein:1988me}
for the $b \to s$ transitions.

The leading order contributions to the Wilson coefficients
$C_{7\gamma,9,10}$ at the order of ${\cal O}(\tan^0\beta)$
in Model A from gluino exchanges are
\begin{eqnarray}
C_{7\gamma} &=& \frac{2}{72 \lambda_b} \frac{g_s^2}{g^2}
   \frac{m_W^2}{m_{\tilde g}^2} \bigg[
   F_{2}(x_{\tilde g}) \delta^{LL}_{12} +
    F^\prime_{2}(x_{\tilde g}) \delta^{LR}_{12} \delta^{LR\ast}_{22}
   - 4 \frac{m_{\tilde g}}{m_c} F_{4}(x_{\tilde g}) \delta^{LR}_{12}
   - 4 \frac{m_{\tilde g}}{m_c} F^\prime_{4}(x_{\tilde g})
    \delta^{LL}_{12} \delta^{LR}_{22} \bigg] ~, \nonumber\\
C_9 &=& \frac{4}{72 \lambda_b} \frac{g_s^2}{g^2}
    \frac{m_W^2}{m_{\tilde g}^2} \bigg[
    f^{\prime}_6(x_{\tilde g}) \delta^{LL}_{12} +
    f^{\prime\prime}_6(x_{\tilde g}) \delta^{LR}_{12} \delta^{LR\ast}_{22}
    \bigg] ~,~\nonumber\\
    &&- \frac{1}{2 \lambda_b s_W^2} \frac{g_s^2}{g^2} (-1 + 4 s_W^2) \bigg[
    - f^{(1)}_{c00}(x_{\tilde g}) \delta^{LR}_{12} \delta^{LR\ast}_{22}
    + f^{(2)}_{c00}(x_{\tilde g}) \delta^{LL}_{12}
    + f^{(3)}_{c00}(x_{\tilde g}) \delta^{LR}_{12} \delta^{LR\ast}_{22}
      \bigg] ~, \nonumber\\
C_{10} &=& - \frac{1}{2 \lambda_b s_W^2} \frac{g_s^2}{g^2} \bigg[
    - f^{(1)}_{c00}(x_{\tilde g}) \delta^{LR}_{12} \delta^{LR\ast}_{22}
    + f^{(2)}_{c00}(x_{\tilde g}) \delta^{LL}_{12}
    + f^{(3)}_{c00}(x_{\tilde g}) \delta^{LR}_{12} \delta^{LR\ast}_{22}
      \bigg] ~,
\end{eqnarray}
where the one-loop functions are defined as follows: 
$f^{\prime}_6(x) = x \frac{\partial f_6(x)}{\partial x}$,
$f^{\prime\prime}_6(x) = \frac{x^2}{2} \frac{\partial^2 f_6(x)}{\partial x^2}$,
$f^{(1)}_{c00}(x) = \frac{x^2}{2}
  \frac{\partial^2 f_{c00}(x,y)}{\partial x \partial y}|_{y->x}$,
$f^{(2)}_{c00}(x) = x \frac{\partial f_{c00}(x,x)}{\partial x}$,
$f^{(3)}_{c00}(x) = \frac{x^2}{2} \frac{\partial^2 f_{c00}(x,x)}{\partial x^2}$,
and $F_{2(4)}$, $F^\prime_{2(4)}$, and $f_{6(c00)}$
are defined in Ref.~\cite{Cheng:2004jf}.

In Model I, the short distance (SD) contribution from
the charged Higgs exchange is negligible,
$B(D^0 \to \gamma \gamma) \sim 10^{-14}$.
In Model II, the contribution can be estimated as
$B(D^0 \to \gamma \gamma) = 2.8 \times 10^{-11}$.
In Model A with a double insertion of
$(\delta^{LL}_{12} \delta^{LR}_{22})$, we have
$C_{7\gamma}^{\tilde g} \sim -2.05 \times
  \frac{(\delta^{LL}_{12} \delta^{LR}_{22})}{10^{-3}}$
from gluino exchange.
The SD contribution can be estimated as
$B(D^0 \to \gamma \gamma) = 5.7 \times 10^{-7}$.
In all three models, we have
$B(D^0 \to \mu^+ \mu^-)$ and $B(D^+ \to \pi^+ \mu^+ \mu^-)$
far below the current experimental bounds.

\section{Conclusion}

We proposed a class of the two-Higgs-doublet SMs with 
a SM singlet and a class of supersymmetric SMs with two pairs of Higgs 
doublets, where the right-handed up/charm quarks and the right-handed 
top quark have different quantum numbers under extra discrete symmetries. 
So the right-handed up and charm quarks couple 
to one Higgs doublet field, while the right-handed top quark couples 
to another Higgs doublet. 
We have studied the direct CP asymmetries in charm hadronic decays
in Models I, II and A.
We found that the large direct CP asymmetry difference
cannot be accommodated within Model I and II with
the contributions of charged Higgs bosons.
In Model A,
we can accommodate the experimental measurement of direct CP asymmetry
with both $O_{8g}$ and $O_{13}$ operators,
while the constraints from the $\Delta c=2$ and
$\Delta c=1$ processes are satisfied.

We leave the detailed studies on phenomenological consequences of our models 
to the future.

\section*{Acknowledgments}

This research was supported in part by the Natural Science Foundation of 
China under Grants No. 10075069, No. 11375248, No. 10821504, No. 11075194,
No. 11135003, No. 11135009, and No. 11275246.

\end{document}